\documentclass[aps,prl,twocolumn,noshowpacs,superscriptaddress,longbibliography]{revtex4}
\usepackage{CJK}
\usepackage{bm}
\usepackage{graphicx}
\usepackage{color}
\usepackage{amsmath,amssymb,amsfonts,amsthm}
\usepackage[colorlinks=true,linkcolor=blue,anchorcolor=red,citecolor=blue,urlcolor=blue]{hyperref}

\begin{document}
\title{Topological Landau Lattice}
\author{Y. X. Zhao}
\email[]{yuxinfruit@gmail.com}
\affiliation{Department of Physics and Center of Theoretical and Computational Physics, The University of Hong Kong, Pokfulam Road, Hong Kong, China}
\affiliation{Max-Planck-Institute for Solid State Research, D-70569 Stuttgart, Germany}
\author{Y. Lu}
\affiliation{Max-Planck-Institute for Solid State Research, D-70569 Stuttgart, Germany}
\author{Hai-Zhou Lu}
\email[]{luhz@sustc.edu.cn}
\affiliation{Institute for Quantum Science and Engineering and Department of Physics,
South University of Science and Technology of China, Shenzhen 518055, China}
\begin{abstract}
The concept of topological fermions, including Weyl and Dirac fermions, stems from the quantum Hall state induced by a magnetic field,
but the definitions and classifications of topological fermions are formulated without using magnetic field. It is unclear whether and how the topological information of topological fermions can be probed once their eigen spectrum is completely rebuilt by a strong magnetic field. In this work, we provide an answer via mapping Landau levels (bands) of topological fermions in $d$ dimensions to the spectrum of a $(d-1)$-dimensional lattice model.
The resultant ``Landau lattice" may correspond to a topological insulator,
and its topological property can be determined by real-space topological invariants.
Accordingly, each zero-energy Landau level (band) inherits the topological stability from the corresponding topological boundary state of the Landau lattice.
The theory is demonstrated in detail by transforming 2D Dirac fermions under magnetic fields to the Su-Schrieffer-Heeger models in class AIII, and 3D Weyl fermions to the Chern insulators in class A.
\end{abstract}

\maketitle

\textit{Introduction--}
The practice of defining phases in quantum matter by their topological properties traces its root back to the quantum Hall effects of electron gases in strong magnetic fields \cite{Klitzing80prl,Thouless82prl}. In the past decades, its generalization in the absence of magnetic field led to the discoveries of new phases of matter such as the topological insulators \cite{Hasan10rmp,Qi11rmp,Shen12book} and semimetals. In particular, classifying and modeling topological massless fermions with various symmetries~\cite{Volovik-book,Horava,ZhaoWang-Classification,Sato-K1,Reflection-SM,Classification-RMP,Zhao-Schnyder-Wang-PT,New-fermions} lie at the frontiers of condensed-matter physics. A number of quantum materials have been proposed and experimentally verified~\cite{Wan11prb,Xu15sci-TaAs,Lv15prx,LuL15sci,Murakami07njp,Wang12prb,Liu14sci,Xu15sci,Burkov11prb,Bian16nc,Schoop16nc,Neupane16prb,Wu16np,WangZJ16nat,New-fermions,ChenW17arXiv,YanZ17arXiv}, among them the topological Weyl~\cite{Wan11prb,Xu15sci-TaAs,Lv15prx,LuL15sci}, Dirac~\cite{Murakami07njp,Wang12prb,Liu14sci,Xu15sci}, and nodal-line~\cite{Burkov11prb,Bian16nc,Schoop16nc,Neupane16prb,Wu16np} semimetals. When exposed to a magnetic field, the topological massless fermions form novel Landau levels (bands) that are essentially different from ordinary Fermi liquids, leading to a number of exotic properties such as the Berry phase corrected half-integer Hall conductance \cite{ZhengYS02prb,Gusynin05prl,Novoselov05nat,ZhangYB05nat,Xu14np,Yoshimi15nc,ZhangSB15srep}, chiral anomaly \cite{Nielsen83plb},
charge density wave \cite{Alicea09prbrc,WangZ13prbrc} and Tomonaga-Luttinger liquid \cite{ZhangXX17prb} with high Landau degeneracy, and Weyl node annihilation \cite{ZhangCL15arXiv-TaP}.
Although one may naturally expect that these phenomena originate from the topological properties of the field-free fermions, a critical question here is whether and how a correspondence can be established between them, knowing that the eigen spectrum of the massless fermions can be completely re-structured in the presence of magnetic fields and the topological information hidden.

In this work, we answer this question by providing a general theory for defining and characterizing the topological properties of the Landau levels, whereby the Landau levels of the topological massless fermions are mapped to
a semi-infinite lattice model, which we refer to as a \textit{Landau lattice}. The Landau lattice can be extended to be infinite, and the field-hidden topological information of massless fermions can be extracted by studying topological properties of the extended Landau lattice.
Because the infinite Landau lattice has no translational symmetry, we employ the topological invariants formulated in real space~\cite{Kitaev-Honeycoumb,Hastings1,Hastings2,Prodan-book}, which has been developed in the context of strong topological insulators and superconductors~\cite{Kane-RMP,XLQi-RMP,Classification-RMP,Schnyder-classification,Kitaev-classification,Ryu-classification}.
Moreover, we discover that the Landau levels (bands) crossing zero energy are topologically protected by the topological configuration of the states with high Landau indices and are stable against perturbations.
As the low and high Landau levels (bands) correspond to the boundary and bulk states in the Landau lattice, respectively, this is a natural consequence of the bulk-boundary correspondence~\cite{Hasugai-Chern,Volovik-book,Kitaev-Honeycoumb,ZhaoWang-BBC,Prodan-book}, which dictates that the topologically protected boundary states are determined by a bulk topological invariant.
The low-energy effective theory for the zero-energy low Landau levels (bands) is then further established by studying the topological invariants of the bulk states in the lattice model.


\textit{Systems in magnetic fields as tight-binding models--}
We start with a 2D system in a perpendicular uniform magnetic field $B$. The gauge covariant momentum operators are 
$\Pi_x=\hat{p}_x-qA_x$ and $\Pi_y=\hat{p}_y-qA_y$ with $B=\partial_x A_y-\partial_y A_x$,
which may be linearly recombined to form the boson creator and annihilator of a harmonic oscillator,
\begin{equation}\label{particle-operators}
a^\dagger=\frac{\ell_B}{\sqrt{2}\hbar}(\Pi_x-i\Pi_y),\ \ \ a=\frac{\ell_B}{\sqrt{2}\hbar}(\Pi_x+i\Pi_y),
\end{equation}
where $\ell_B=\sqrt{\hbar/ qB}$ and $[a,a^\dagger]=1$. The operators act on the Hilbert space spanned by the orthonormal particle-number basis $|n\rangle=(a^\dagger)^n|0\rangle/\sqrt{n!}$ with integers $n=0,1,2,\cdots$.
The creator adds a particle and the annihilator destructs one, which are implemented by
\begin{equation}
\begin{split}
& a|0\rangle=0, \quad a|n\rangle=\sqrt{n}|n-1\rangle,~n>0,\\
& a^\dagger|n\rangle=\sqrt{n+1}|n+1\rangle,~~n=0,1,2,\cdots.
\end{split}
\end{equation}
The eigenvalue of the particle number $N=a^\dagger a$ is then the Landau index $n$.

\begin{figure}
	\includegraphics[scale=1]{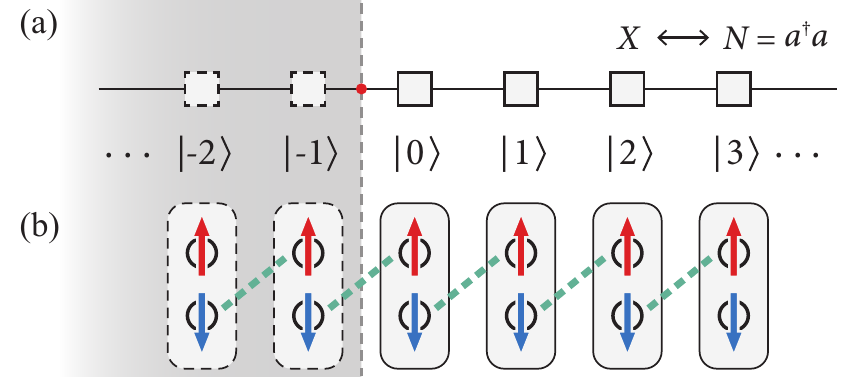}
	\caption{(a)~Identification of the Hilbert space of a quantum oscillator to the spatial Hilbert space of a semi-infinite lattice, where the particle number $N$ is equivalent to the site position $X$. The semi-infinite lattice is then extended to be infinite for the purpose of studying its topological properties. (b)Pairings of states in a lattice model derived from Dirac fermions in a magnetic field. This is a reminiscence of the dimerization ocurring in the well-known SSH model. \label{Nearest-Pairing}}
\end{figure}

We then construct the real-space component of the Hilbert space for a tight-binding model on a 1D semi-infinite lattice. The lattice sites are labelled as $n=0,1,2,\cdots$, each assigned with a quantum state $|n\rangle$, as illustrated in Fig.~\ref{Nearest-Pairing}(a). As a 1D lattice, the forward and backward translators, denoted by $\widehat{S}^\dagger$ and $\widehat{S}$, respectively, act on the Hilbert space as
\begin{eqnarray}
&& \widehat{S}|0\rangle=0,\quad \widehat{S}|n\rangle=|n-1\rangle,~~n=1,2,\cdots,\\
&& \widehat{S}^\dagger|n\rangle=|n+1\rangle ,~~n=0,1,2,\cdots.
\end{eqnarray}
Another useful operator for the lattice is the position operator $X$ which reads off the position of a state,
\begin{equation}
X|n\rangle=n|n\rangle,~~n=0,1,2,\cdots.
\end{equation}
It is noteworthy that $\widehat{S}$ and $\hat{S}^\dagger$ commute for every state except the end state of the semi-infinite lattice $|0\rangle$, namely
\begin{equation}
[\widehat{S},\widehat{S}^\dagger]=\delta_{0,X}.
\end{equation}
We then identify the real-space Hilbert space of the semi-infinite 1D lattice with that of the harmonic oscillator. In this identification, the position operator $X$ is just the Landau index operator $N=a^\dagger a$, and the translators can be expressed as
$
\widehat{S}^\dagger=a^\dagger \frac{1}{\sqrt{N+1}}$, $ \widehat{S}=\frac{1}{\sqrt{N+1}} a
$. Inversely, with $N=X$,
\begin{equation}
a^\dagger=\widehat{S}^\dagger \sqrt{X+1},\quad a=\sqrt{X+1}\,\widehat{S}. \label{mappings}
\end{equation}
Before proceeding, we note that topological configurations shall be defined for the entirety of Landau levels, namely in the quotient space by the degeneracy of each Landau level. This should be distinguished from the quantum hall effects, where the topology is defined in the degenerate subspaces.

In general, a coarse-grained Hamiltonian for topological fermions in a condensed-matter system may be cast into the form
$\mathcal{H}(\mathbf{p})=\sum_a f_a(\mathbf{p})\Gamma^a$, where $\Gamma^a$ are a set of matrices, Dirac matrices for instance, accounting for the internal degrees of freedom for topological fermions, and $\mathbf{p}$ is the effective momentum in a $d$-dimensional momentum space. Such a Hamiltonian may describe low-energy excitations arising from band crossings in a band structure. If a field $B$ is applied normal to the $x_j$-$x_k$ plane, the corresponding gauge potentials satisfy $B=\partial_j A_k-\partial_k A_j $,
and the momenta $p_j$ and $p_k$ are gauged to be $\Pi_j=p_j-qA_j$ and $\Pi_k=p_k-qA_k$, respectively. The recombined operators $a^\dagger$ and $a$ are mapped to be forward and backward translators $\widehat{S}^\dagger$ and $\widehat{S}$ in a 1D lattice, whose sites are labelled by $X$. By such construction, a $d$D model of topological fermions under a constant field $B$ is transformed to a $(d-1)$D semi-infinite lattice with the $(d-2)$D boundary perpendicular to $X$, which we refer to as a \textit{Landau lattice} of the topological fermions. More generally, for $d\le 2n$, if $n$ constant fields are applied, we obtain a $(d-n)$D lattice with $n$ boundaries perpendicular to the $n$ independent dimensions via the transformation. For physical dimensions $d=3$ and $2$, a magnetic field reduces the dimensionality to $2$ and $1$, respectively.

We then extend the semi-infinite Landau lattice to be infinite by adding states $|-1\rangle,~|-2\rangle,~\cdots$, as illustrated in Fig.~\ref{Nearest-Pairing}(a). On the infinite Landau lattice, the two translators are denoted by ``removing hats'', and accordingly act as
\begin{equation}
S^\dagger|n\rangle=|n+1\rangle,\quad S|n\rangle=|n-1\rangle,
\end{equation}
with $n=\cdots,-2,-1,0,1,2,\cdots$. The extended operators are invertible and mutually independent, satisfying the relations
\begin{equation}
S^\dagger S=1,\quad [S^\dagger,S]=0.
\end{equation}


We now have a $(d-1)$D infinite lattice model---the extended Landau lattice---derived from $d$D topological fermions under a uniform magnetic field. The infinite lattice model may be a topological insulator or (semi)metal that hosts topological gapless boundary modes when a $(d-2)$D boundary is opened perpendicular to the $X$ direction. Inversely, these topological boundary modes correspond to the Landau levels (bands) of fermions that are of small Landau indices, and are located at (go across) zero energy.
Since the boundary modes of the semi-infinite lattice are topologically protected by the bulk topological invariants, these corresponding Landau levels of small indices are topologically protected by nontrivial topological configurations encoded deep into the Landau levels of large indices. Similar to the robustness of the topological boundary modes, these corresponding topological Landau levels (bands) with small Landau indices are also stable against (symmetry-preserving) disorder and perturbations. It is noteworthy that the resultant infinite lattice model has no translational symmetry along the $X$-direction due to the position-dependent mappings in Eq.~\eqref{mappings}. Thus, in contrast to the common practice of using topological invariants in terms of momentum-space Hamiltonian or Berry connection in the presence of translational symmetry, it is required here to employ topological invariants directly formulated in real space,

\textit{2D Dirac fermions and the SSH models--}
In the following, we proceed to illustrate and demonstrate the somewhat abstract general theory by elementary examples, which are interesting in their own right.
Let us first consider 2D Dirac fermions with the Hamiltonian
\begin{equation}
\mathcal{H}_D=\frac{v}{2}(p_-\sigma^++p_+\sigma^-),
\end{equation}
where $p_{\pm}=p_x\pm i p_y$ and $\sigma^{\pm}=\sigma_1\pm i\sigma_2$. Here $\sigma_j$ are the Pauli matrices.
Coupling Dirac fermions to the magnetic field is realized by substitutions
$p_+\rightarrow \sqrt{2}\hbar a/\ell_B$ and $p_-\rightarrow \sqrt{2}\hbar a^\dagger/\ell_B$,
which lead to
$\mathcal{H}_D^B=(\hbar\omega_B/2)(a^\dagger \sigma^++a\sigma^-)$
with $\omega_B=\sqrt{2}v/\ell_B$. 
The corresponding Landau lattice can be derived using Eq.~\eqref{mappings} as
$
\widehat{\mathcal{H}}_{L}=(\hbar \omega_B/2)(\widehat{S}^\dagger \sqrt{X+1}\sigma^++\sqrt{X+1}\widehat{S}\sigma^-)
$,
which inherits the chiral symmetry $\{\widehat{\mathcal{H}}_{L},\sigma_3\}=0$ and is therefore in class AIII. The chiral symmetry $\sigma_3$ charges a state $|\psi\rangle$ with positive or negative chirality, i.e., $\sigma_3|\psi\rangle=\pm|\psi\rangle$.
The obtained lattice model has an energy gap in the bulk, and one can determine whether it has topologically protected zero end-states by extending it to an infinite lattice. Note that the extension is not unique, and the inversion-symmetric one is given as
\begin{equation}
\mathcal{H}_{L} =\frac{\hbar \omega_B}{2}(S^\dagger \sqrt{|X|+1}~\sigma^++\sqrt{|X|+1}~S~\sigma^-) \label{Inft-AIII-model}.
\end{equation}

Let us recall some basics of topological insulators in class AIII. The Hamiltonian can be written in a general form
\begin{equation}
\mathcal{H}_{L}=\begin{pmatrix}
0 & Q^\dagger\\
Q & 0
\end{pmatrix}
\end{equation}
with the chiral symmetry $\sigma_3$, and the topological invariant for the $\mathbb{Z}$ classification can be given in real space as
\begin{equation}
\nu=\lim_{\mathcal{N}\rightarrow\infty}\frac{1}{\mathcal{N}}\mathrm{Tr}^\prime~Q^{-1}[Q,X],\label{chiral-invariant}
\end{equation}
where $\mathcal{N}$ is the number of sites~\cite{Kitaev-Honeycoumb,Notes-topological-number}, which when sent to infinity quantizes the topological invariant $\nu$ to an integer. A positive (negative) $\nu$ corresponds to $\nu$ (-$\nu$) zero-modes with positive (negative) chirality concentrated at the end of the semi-infinite lattice. Note that the topological invariant in real space requires no translation invariance and thus serves well for our purpose.

\begin{figure}
	\includegraphics[scale=1.0]{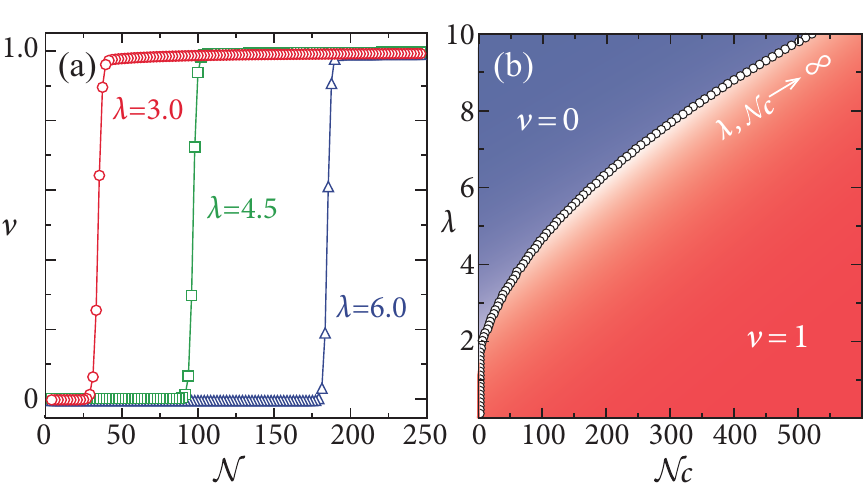}
	\caption{(a)~Topological invariant $\nu$ with increasing number of lattice sites $\mathcal{N}$ for different $\lambda$'s. (b)~Dependence of critical site number $\mathcal{N}_c$ to $\lambda$. As a result of data fitting, $\mathcal{N}_c=5.5 \lambda^2-1.3 \lambda-14$. \label{Numeric-Invariant} }
\end{figure}

For the case of Eq.~\eqref{Inft-AIII-model} with $Q=\hbar \omega_B \sqrt{|X|+1}S$, the topological invariant can then be directly computed by Eq.~\eqref{chiral-invariant}. As shown in Fig.~\ref{Numeric-Invariant}, the topological invariant converges to $1$ with increasing $\mathcal{N}$. Alternatively, one may reduce this case to the simplest topologically nontrivial Su-Schrieffer-Heeger (SSH) model~\cite{SSH-Model-1,SSH-Model-2} with $Q=S$ by simplifying the topological invariant to $\nu=\lim_{\mathcal{N}\rightarrow\infty}\frac{1}{\mathcal{N}}\mathrm{Tr}^\prime~S^\dagger[X,S]$, or through a continuous deformation $Q(\tau)=\hbar\omega_B\sqrt{|X|(1-\tau)+1}S$ with $\tau$ varying from $0$ to $1$, in course of which the bulk gap is always open and the topological invariant unchanged.

According to the bulk-boundary correspondence, there is a topologically-protected zero mode residing at the end of the semi-infinite lattice. Translating this back to the field-coupled Dirac fermions, this implies that there exists a Landau level composed of states with small Landau indices. We find the zero-energy state to be $|0\rangle\otimes|\uparrow\rangle$,
which has positive chirality, in accord with the bulk topological invariant.
The rest of the Landau levels can be interpreted as bulk states, each of which is a superposition of states from nearest neighbors on the lattice. In other words, they correspond to the bulk states of the SSH model with the pairings of the bulk states, $|n\rangle\otimes |\downarrow\rangle$ and $|n+1\rangle\otimes |\uparrow\rangle$, where $n=0,1,2,\cdots$,
and the Hamiltonian is block-diagonalized in the subspace of each pairing, as illustrated in Fig.~\ref{Nearest-Pairing}. Further diagonalizing the Hamiltonian in each block solves the spectrum $E_{n}^{\pm}=\pm\hbar\omega_B\sqrt{n+1}$ of the eigenstates $(1/\sqrt{2}) ( |n,\downarrow\rangle\pm |n+1,\uparrow\rangle~)$ with $n=0,1,2,\cdots$.

Now we elaborate a significant difference in phase transition between the Landau lattice model in class AIII and the translation-invariant SSH model. Let us recall the generic SSH model with a symmetry-preserving on-site term, $\mathcal{H}_{SSH}$$=$$ (S^\dagger\sigma^+$$+$$S\sigma^-)/2$$+$$\lambda\sigma_2$, which corresponds to $\widetilde{Q}=S+i\lambda$. For $|\lambda|>1$, the model is in a trivial phase without end states since the on-site term is dominant. On the other hand, an arbitrarily large $\lambda$ cannot demote the Landau lattice with $Q=\sqrt{|X|+1}S+i\lambda$ to a trivial phase. As shown by numerical calculations in Fig.~\ref{Numeric-Invariant}, $\nu$ always approaches $1$ for any finite $\lambda$ when the lattice length $\mathcal{N}$ is sufficiently large. More precisely, as seen from Fig.~\ref{Numeric-Invariant}(a), there exists a critical site number $\mathcal{N}_c$, exceeding which the topological invariant jumps from $0$ to $1$. Numerical fitting of the results in Fig.~\ref{Numeric-Invariant}(b) shows that $\mathcal{N}_c$ is approximately a quadratic function of $\lambda$, whose explicit form reads $\mathcal{N}_c=5.5 \lambda^2-1.3 \lambda-14$. We emphasize that the criticality here corresponds to no phase transition in the ordinary sense, since the notion of a phase is defined under the thermodynamic limit $\mathcal{N}\rightarrow \infty$. The fact that the Landau lattice is always topologically nontrivial can be intuitively understood as follows: The nearest hopping term $\sqrt{|X|+1}S$ responsible for the topology will always dominate with increasing magnitude when going deep enough into the bulk. What actually happens with increasing $|\lambda|$ is that the concentration of the end state $|\psi\rangle$ is moved towards the bulk, and its spatial distribution is smoothed [see the Supplemental Material (SM)~\cite{Supp} for details]. But it is inescapable to be bounded to the end for a finite $\lambda$ on the semi-infinite lattice, since $\langle n,\sigma|\psi\rangle$ always approaches $0$ as $n\rightarrow \infty$. Accordingly, under an arbitrarily large perturbation $\lambda\sigma_2$, there is always a zero-energy Landau level that is a superposition dominated by states with small Landau indices.

Recalling that 1D topological insulators in class AIII have an integer classification, we can construct the corresponding Landau levels of topological fermions with an arbitrary number of topologically protected zero-modes mainly composed of small indices, which can be found in SM~\cite{Supp}.

\textit{Weyl fermions and Chern insulators--}
We now consider a 3D example in class A without any symmetry, namely the Weyl-fermion Hamiltonian,
$\mathcal{H}_W^{L/R}(\mathbf{p})=p_x\sigma_1+p_y\sigma_2\pm p_z\sigma_3$,
whose infinite Landau lattice is derived as
\begin{equation}
\mathcal{H}_{W,L}^{L/R}=\frac{1}{2}(\sqrt{|X|+1}S\sigma^-+\mathrm{h.c.})\pm k_z\sigma_3,
\end{equation}
where the $x$-direction is expressed in real space without translational symmetry, while the $z$-direction remains in momentum space with translational symmetry.
This is a 2D lattice with an energy gap, since $\sqrt{X+1}S$ is invertible, and $\sigma_3$ anticommutes with $\sigma^{\pm}$. The model can be continuously deformed to be $\widetilde{\mathcal{H}}_{W,L}^{\pm}=S^\dagger\sigma^+ + S\sigma^-\pm k_z\sigma_3$ without closing the gap,
so that translational symmetry along the $x$-direction is acquired, entirely parallel to what we did for Eq.~\eqref{Inft-AIII-model}. Now the model may be a Chern insulator~\cite{Haldane-Model,TKNN}, and the Chern number can be calculated using translational symmetries, although a formula of the Chern number in real space is also available~\cite{Kitaev-Honeycoumb}. In momentum space, the Hamiltonian is
$\widetilde{\mathcal{H}}_{W,L}^{L/R}(k_x,k_z)=e^{ik_xx} \sigma^+ + e^{-ik_xx} \sigma^-\pm k_z\sigma_3$,
and the Chern number is just the unit winding number of the vector field $\mathbf{d}=(\cos k_x,\sin k_x,\pm k_z)$~\cite{Volovik-book}, after normalized to be $\hat{\mathbf{d}}$, as a mapping from $S^2$ to $S^2$. It can be computed explicitly as $\nu=(1/4\pi)\int_{-\pi}^{\pi} dk_x\int_{-\infty}^{\infty} dk_z~\hat{\mathbf{d}}\cdot(\partial_{k_z} \hat{\mathbf{d}}\times \partial_{k_x} \hat{\mathbf{d}})$.
The positive (negative) unit Chern number implies a right (left)-moving edge band across zero energy for any 1D edge of a semi-infinite system. In this particular case where all sites left to $X=0$ are cut off, making an edge for the $x$-dimension, the unit Chern number corresponds to a chiral band parametrized by $k_z$, which are gapless edge modes for the semi-infinite lattice model, or equivalently a chiral Landau band dominated by states with small indices (see the SM~\cite{Supp} for a graphic illustration). The existence of the chiral Landau band is topologically protected against weak disorder and perturbations. Chern insulators of higher Chern numbers correspond to Landau bands of multiply-charged Weyl fermions, which is elaborated in the SM similar to the previous case of symmetry class AIII~\cite{Supp}.

In a two-band minimal model of Weyl semimetals, for instance $\mathcal{H}_{WSM}=p_x\sigma_1+p_y\sigma_2-(p_z^2-\mu^2)\sigma_3$, there are two Weyl points each with left and right chirality, conforming with the Nielsen-Ninomiya no-go theorem~\cite{NN-NoGo,ZhaoWang-NoGo}. By applying a uniform magnetic field along the $z$-direction, according to the discussions above, the right (left) Weyl point leads to a right (left) massless mover, which is concentrated at the smallest Landau index. In the virtual Landau band structure as graphically illustrated in the SM~\cite{Supp}, the left and right moving modes should be connected smoothly, leading to a band parametrized by $k_z$ as a reminiscence of the Tomonaga-Luttinger liquid when all negative states are filled by electrons. Again such topological properties are insensitive to weak disorder and perturbations.

\textit{Summary} -
In summary, we bridge the Landau levels (bands) of $d$-dimensional topological fermions to a $(d-1)$D Landau lattice with $d\ge 2$. It is shown that the topological boundary modes of the Landau lattice, resulting from the bulk topological invariant, correspond to topological low Landau levels (bands) crossing zero energy. Thus, these zero modes among Landau levels (bands) are protected by topological configurations deep into the gapped states with large Landau indices, which are therefore robust against (symmetry-preserving) weak disorder and perturbations.

\textit{Acknowledgements} -
This work was supported by the National Key R \& D Program (Grant
No. 2016YFA0301700), National Natural Science Foun-
dation of China (Grant No. 11574127),
and Guangdong Innovative and Entrepreneurial Research
Team Program (Grant No. 2016ZT06D348).

\bibliographystyle{apsrev}
\bibliography{Landau-SSH,refs-transport}

\begin{thebibliography}{62}
\expandafter\ifx\csname natexlab\endcsname\relax\def\natexlab#1{#1}\fi
\expandafter\ifx\csname bibnamefont\endcsname\relax
  \def\bibnamefont#1{#1}\fi
\expandafter\ifx\csname bibfnamefont\endcsname\relax
  \def\bibfnamefont#1{#1}\fi
\expandafter\ifx\csname citenamefont\endcsname\relax
  \def\citenamefont#1{#1}\fi
\expandafter\ifx\csname url\endcsname\relax
  \def\url#1{\texttt{#1}}\fi
\expandafter\ifx\csname urlprefix\endcsname\relax\def\urlprefix{URL }\fi
\providecommand{\bibinfo}[2]{#2}
\providecommand{\eprint}[2][]{\url{#2}}

\bibitem[{\citenamefont{Klitzing et~al.}(1980)\citenamefont{Klitzing, Dorda,
  and Pepper}}]{Klitzing80prl}
\bibinfo{author}{\bibfnamefont{K.~v.} \bibnamefont{Klitzing}},
  \bibinfo{author}{\bibfnamefont{G.}~\bibnamefont{Dorda}}, \bibnamefont{and}
  \bibinfo{author}{\bibfnamefont{M.}~\bibnamefont{Pepper}},
  \bibinfo{journal}{Phys. Rev. Lett.} \textbf{\bibinfo{volume}{45}},
  \bibinfo{pages}{494} (\bibinfo{year}{1980}),
  \urlprefix\url{http://link.aps.org/doi/10.1103/PhysRevLett.45.494}.

\bibitem[{\citenamefont{Thouless
  et~al.}(1982{\natexlab{a}})\citenamefont{Thouless, Kohmoto, Nightingale, and
  den Nijs}}]{Thouless82prl}
\bibinfo{author}{\bibfnamefont{D.~J.} \bibnamefont{Thouless}},
  \bibinfo{author}{\bibfnamefont{M.}~\bibnamefont{Kohmoto}},
  \bibinfo{author}{\bibfnamefont{M.~P.} \bibnamefont{Nightingale}},
  \bibnamefont{and} \bibinfo{author}{\bibfnamefont{M.}~\bibnamefont{den Nijs}},
  \bibinfo{journal}{Phys. Rev. Lett.} \textbf{\bibinfo{volume}{49}},
  \bibinfo{pages}{405} (\bibinfo{year}{1982}{\natexlab{a}}),
  \urlprefix\url{http://link.aps.org/doi/10.1103/PhysRevLett.49.405}.

\bibitem[{\citenamefont{Hasan and Kane}(2010{\natexlab{a}})}]{Hasan10rmp}
\bibinfo{author}{\bibfnamefont{M.~Z.} \bibnamefont{Hasan}} \bibnamefont{and}
  \bibinfo{author}{\bibfnamefont{C.~L.} \bibnamefont{Kane}},
  \bibinfo{journal}{Rev. Mod. Phys.} \textbf{\bibinfo{volume}{82}},
  \bibinfo{pages}{3045} (\bibinfo{year}{2010}{\natexlab{a}}),
  \urlprefix\url{http://link.aps.org/doi/10.1103/RevModPhys.82.3045}.

\bibitem[{\citenamefont{Qi and Zhang}(2011{\natexlab{a}})}]{Qi11rmp}
\bibinfo{author}{\bibfnamefont{X.-L.} \bibnamefont{Qi}} \bibnamefont{and}
  \bibinfo{author}{\bibfnamefont{S.-C.} \bibnamefont{Zhang}},
  \bibinfo{journal}{Rev. Mod. Phys.} \textbf{\bibinfo{volume}{83}},
  \bibinfo{pages}{1057} (\bibinfo{year}{2011}{\natexlab{a}}),
  \urlprefix\url{http://link.aps.org/doi/10.1103/RevModPhys.83.1057}.

\bibitem[{\citenamefont{Shen}(2012)}]{Shen12book}
\bibinfo{author}{\bibfnamefont{S.-Q.} \bibnamefont{Shen}},
  \emph{\bibinfo{title}{Topological Insulators}}
  (\bibinfo{publisher}{Springer-Verlag}, \bibinfo{address}{Berlin Heidelberg},
  \bibinfo{year}{2012}).

\bibitem[{\citenamefont{Volovik}(2003)}]{Volovik-book}
\bibinfo{author}{\bibfnamefont{G.~E.} \bibnamefont{Volovik}},
  \emph{\bibinfo{title}{Universe in a helium droplet}}
  (\bibinfo{publisher}{Oxford University Press, Oxford UK},
  \bibinfo{year}{2003}), ISBN \bibinfo{isbn}{0521670535}.

\bibitem[{\citenamefont{Ho\ifmmode~\check{r}\else
  \v{r}\fi{}ava}(2005)}]{Horava}
\bibinfo{author}{\bibfnamefont{P.}~\bibnamefont{Ho\ifmmode~\check{r}\else
  \v{r}\fi{}ava}}, \bibinfo{journal}{Phys. Rev. Lett.}
  \textbf{\bibinfo{volume}{95}}, \bibinfo{pages}{016405}
  (\bibinfo{year}{2005}),
  \urlprefix\url{https://link.aps.org/doi/10.1103/PhysRevLett.95.016405}.

\bibitem[{\citenamefont{Zhao and Wang}(2013)}]{ZhaoWang-Classification}
\bibinfo{author}{\bibfnamefont{Y.~X.} \bibnamefont{Zhao}} \bibnamefont{and}
  \bibinfo{author}{\bibfnamefont{Z.~D.} \bibnamefont{Wang}},
  \bibinfo{journal}{Phys. Rev. Lett.} \textbf{\bibinfo{volume}{110}},
  \bibinfo{pages}{240404} (\bibinfo{year}{2013}),
  \urlprefix\url{http://link.aps.org/doi/10.1103/PhysRevLett.110.240404}.

\bibitem[{\citenamefont{Shiozaki and Sato}(2014)}]{Sato-K1}
\bibinfo{author}{\bibfnamefont{K.}~\bibnamefont{Shiozaki}} \bibnamefont{and}
  \bibinfo{author}{\bibfnamefont{M.}~\bibnamefont{Sato}},
  \bibinfo{journal}{Phys. Rev. B} \textbf{\bibinfo{volume}{90}},
  \bibinfo{pages}{165114} (\bibinfo{year}{2014}),
  \urlprefix\url{http://link.aps.org/doi/10.1103/PhysRevB.90.165114}.

\bibitem[{\citenamefont{Chiu and Schnyder}(2014)}]{Reflection-SM}
\bibinfo{author}{\bibfnamefont{C.-K.} \bibnamefont{Chiu}} \bibnamefont{and}
  \bibinfo{author}{\bibfnamefont{A.~P.} \bibnamefont{Schnyder}},
  \bibinfo{journal}{Phys. Rev. B} \textbf{\bibinfo{volume}{90}},
  \bibinfo{pages}{205136} (\bibinfo{year}{2014}),
  \urlprefix\url{http://link.aps.org/doi/10.1103/PhysRevB.90.205136}.

\bibitem[{\citenamefont{Chiu et~al.}(2016)\citenamefont{Chiu, Teo, Schnyder,
  and Ryu}}]{Classification-RMP}
\bibinfo{author}{\bibfnamefont{C.-K.} \bibnamefont{Chiu}},
  \bibinfo{author}{\bibfnamefont{J.~C.~Y.} \bibnamefont{Teo}},
  \bibinfo{author}{\bibfnamefont{A.~P.} \bibnamefont{Schnyder}},
  \bibnamefont{and} \bibinfo{author}{\bibfnamefont{S.}~\bibnamefont{Ryu}},
  \bibinfo{journal}{Rev. Mod. Phys.} \textbf{\bibinfo{volume}{88}},
  \bibinfo{pages}{035005} (\bibinfo{year}{2016}),
  \urlprefix\url{http://link.aps.org/doi/10.1103/RevModPhys.88.035005}.

\bibitem[{\citenamefont{Zhao et~al.}(2016)\citenamefont{Zhao, Schnyder, and
  Wang}}]{Zhao-Schnyder-Wang-PT}
\bibinfo{author}{\bibfnamefont{Y.~X.} \bibnamefont{Zhao}},
  \bibinfo{author}{\bibfnamefont{A.~P.} \bibnamefont{Schnyder}},
  \bibnamefont{and} \bibinfo{author}{\bibfnamefont{Z.~D.} \bibnamefont{Wang}},
  \bibinfo{journal}{Phys. Rev. Lett.} \textbf{\bibinfo{volume}{116}},
  \bibinfo{pages}{156402} (\bibinfo{year}{2016}),
  \urlprefix\url{http://link.aps.org/doi/10.1103/PhysRevLett.116.156402}.

\bibitem[{\citenamefont{Bradlyn et~al.}(2016)\citenamefont{Bradlyn, Cano, Wang,
  Vergniory, Felser, Cava, and Bernevig}}]{New-fermions}
\bibinfo{author}{\bibfnamefont{B.}~\bibnamefont{Bradlyn}},
  \bibinfo{author}{\bibfnamefont{J.}~\bibnamefont{Cano}},
  \bibinfo{author}{\bibfnamefont{Z.}~\bibnamefont{Wang}},
  \bibinfo{author}{\bibfnamefont{M.~G.} \bibnamefont{Vergniory}},
  \bibinfo{author}{\bibfnamefont{C.}~\bibnamefont{Felser}},
  \bibinfo{author}{\bibfnamefont{R.~J.} \bibnamefont{Cava}}, \bibnamefont{and}
  \bibinfo{author}{\bibfnamefont{B.~A.} \bibnamefont{Bernevig}},
  \bibinfo{journal}{Science} \textbf{\bibinfo{volume}{353}}
  (\bibinfo{year}{2016}),
  \urlprefix\url{http://science.sciencemag.org/content/early/2016/07/20/science.aaf5037}.

\bibitem[{\citenamefont{Wan et~al.}(2011)\citenamefont{Wan, Turner, Vishwanath,
  and Savrasov}}]{Wan11prb}
\bibinfo{author}{\bibfnamefont{X.}~\bibnamefont{Wan}},
  \bibinfo{author}{\bibfnamefont{A.~M.} \bibnamefont{Turner}},
  \bibinfo{author}{\bibfnamefont{A.}~\bibnamefont{Vishwanath}},
  \bibnamefont{and} \bibinfo{author}{\bibfnamefont{S.~Y.}
  \bibnamefont{Savrasov}}, \bibinfo{journal}{Phys. Rev. B}
  \textbf{\bibinfo{volume}{83}}, \bibinfo{pages}{205101}
  (\bibinfo{year}{2011}),
  \urlprefix\url{http://link.aps.org/doi/10.1103/PhysRevB.83.205101}.

\bibitem[{\citenamefont{Xu et~al.}(2015{\natexlab{a}})\citenamefont{Xu,
  Belopolski, Alidoust, Neupane, Bian, Zhang, Sankar, Chang, Yuan, Lee
  et~al.}}]{Xu15sci-TaAs}
\bibinfo{author}{\bibfnamefont{S.~Y.} \bibnamefont{Xu}},
  \bibinfo{author}{\bibfnamefont{I.}~\bibnamefont{Belopolski}},
  \bibinfo{author}{\bibfnamefont{N.}~\bibnamefont{Alidoust}},
  \bibinfo{author}{\bibfnamefont{M.}~\bibnamefont{Neupane}},
  \bibinfo{author}{\bibfnamefont{G.}~\bibnamefont{Bian}},
  \bibinfo{author}{\bibfnamefont{C.~L.} \bibnamefont{Zhang}},
  \bibinfo{author}{\bibfnamefont{R.}~\bibnamefont{Sankar}},
  \bibinfo{author}{\bibfnamefont{G.~Q.} \bibnamefont{Chang}},
  \bibinfo{author}{\bibfnamefont{Z.~J.} \bibnamefont{Yuan}},
  \bibinfo{author}{\bibfnamefont{C.~C.} \bibnamefont{Lee}},
  \bibnamefont{et~al.}, \bibinfo{journal}{Science}
  \textbf{\bibinfo{volume}{349}}, \bibinfo{pages}{613}
  (\bibinfo{year}{2015}{\natexlab{a}}),
  \urlprefix\url{http://www.sciencemag.org/content/349/6248/613.abstract}.

\bibitem[{\citenamefont{Lv et~al.}(2015)\citenamefont{Lv, Weng, Fu, Wang, Miao,
  Ma, Richard, Huang, Zhao, Chen et~al.}}]{Lv15prx}
\bibinfo{author}{\bibfnamefont{B.~Q.} \bibnamefont{Lv}},
  \bibinfo{author}{\bibfnamefont{H.~M.} \bibnamefont{Weng}},
  \bibinfo{author}{\bibfnamefont{B.~B.} \bibnamefont{Fu}},
  \bibinfo{author}{\bibfnamefont{X.~P.} \bibnamefont{Wang}},
  \bibinfo{author}{\bibfnamefont{H.}~\bibnamefont{Miao}},
  \bibinfo{author}{\bibfnamefont{J.}~\bibnamefont{Ma}},
  \bibinfo{author}{\bibfnamefont{P.}~\bibnamefont{Richard}},
  \bibinfo{author}{\bibfnamefont{X.~C.} \bibnamefont{Huang}},
  \bibinfo{author}{\bibfnamefont{L.~X.} \bibnamefont{Zhao}},
  \bibinfo{author}{\bibfnamefont{G.~F.} \bibnamefont{Chen}},
  \bibnamefont{et~al.}, \bibinfo{journal}{Phys. Rev. X}
  \textbf{\bibinfo{volume}{5}}, \bibinfo{pages}{031013} (\bibinfo{year}{2015}),
  \urlprefix\url{http://link.aps.org/doi/10.1103/PhysRevX.5.031013}.

\bibitem[{\citenamefont{Lu et~al.}(2015)\citenamefont{Lu, Wang, Ye, Ran, Fu,
  Joannopoulos, and Solja{\v c}i{\'c}}}]{LuL15sci}
\bibinfo{author}{\bibfnamefont{L.}~\bibnamefont{Lu}},
  \bibinfo{author}{\bibfnamefont{Z.}~\bibnamefont{Wang}},
  \bibinfo{author}{\bibfnamefont{D.}~\bibnamefont{Ye}},
  \bibinfo{author}{\bibfnamefont{L.}~\bibnamefont{Ran}},
  \bibinfo{author}{\bibfnamefont{L.}~\bibnamefont{Fu}},
  \bibinfo{author}{\bibfnamefont{J.~D.} \bibnamefont{Joannopoulos}},
  \bibnamefont{and} \bibinfo{author}{\bibfnamefont{M.}~\bibnamefont{Solja{\v
  c}i{\'c}}}, \bibinfo{journal}{Science} \textbf{\bibinfo{volume}{349}},
  \bibinfo{pages}{622} (\bibinfo{year}{2015}), ISSN \bibinfo{issn}{0036-8075}.

\bibitem[{\citenamefont{Murakami}(2007)}]{Murakami07njp}
\bibinfo{author}{\bibfnamefont{S.}~\bibnamefont{Murakami}},
  \bibinfo{journal}{New. J. Phys.} \textbf{\bibinfo{volume}{9}},
  \bibinfo{pages}{356} (\bibinfo{year}{2007}).

\bibitem[{\citenamefont{Wang et~al.}(2012)\citenamefont{Wang, Sun, Chen,
  Franchini, Xu, Weng, Dai, and Fang}}]{Wang12prb}
\bibinfo{author}{\bibfnamefont{Z.}~\bibnamefont{Wang}},
  \bibinfo{author}{\bibfnamefont{Y.}~\bibnamefont{Sun}},
  \bibinfo{author}{\bibfnamefont{X.~Q.} \bibnamefont{Chen}},
  \bibinfo{author}{\bibfnamefont{C.}~\bibnamefont{Franchini}},
  \bibinfo{author}{\bibfnamefont{G.}~\bibnamefont{Xu}},
  \bibinfo{author}{\bibfnamefont{H.}~\bibnamefont{Weng}},
  \bibinfo{author}{\bibfnamefont{X.}~\bibnamefont{Dai}}, \bibnamefont{and}
  \bibinfo{author}{\bibfnamefont{Z.}~\bibnamefont{Fang}},
  \bibinfo{journal}{Phys. Rev. B} \textbf{\bibinfo{volume}{85}},
  \bibinfo{pages}{195320} (\bibinfo{year}{2012}),
  \urlprefix\url{http://link.aps.org/doi/10.1103/PhysRevB.85.195320}.

\bibitem[{\citenamefont{Liu et~al.}(2014)\citenamefont{Liu, Zhou, Zhang, Wang,
  Weng, Prabhakaran, Mo, Shen, Fang, Dai et~al.}}]{Liu14sci}
\bibinfo{author}{\bibfnamefont{Z.~K.} \bibnamefont{Liu}},
  \bibinfo{author}{\bibfnamefont{B.}~\bibnamefont{Zhou}},
  \bibinfo{author}{\bibfnamefont{Y.}~\bibnamefont{Zhang}},
  \bibinfo{author}{\bibfnamefont{Z.~J.} \bibnamefont{Wang}},
  \bibinfo{author}{\bibfnamefont{H.~M.} \bibnamefont{Weng}},
  \bibinfo{author}{\bibfnamefont{D.}~\bibnamefont{Prabhakaran}},
  \bibinfo{author}{\bibfnamefont{S.~K.} \bibnamefont{Mo}},
  \bibinfo{author}{\bibfnamefont{Z.~X.} \bibnamefont{Shen}},
  \bibinfo{author}{\bibfnamefont{Z.}~\bibnamefont{Fang}},
  \bibinfo{author}{\bibfnamefont{X.}~\bibnamefont{Dai}}, \bibnamefont{et~al.},
  \bibinfo{journal}{Science} \textbf{\bibinfo{volume}{343}},
  \bibinfo{pages}{864} (\bibinfo{year}{2014}),
  \urlprefix\url{http://www.sciencemag.org/content/343/6173/864.abstract}.

\bibitem[{\citenamefont{Xu et~al.}(2015{\natexlab{b}})\citenamefont{Xu, Liu,
  Kushwaha, Sankar, Krizan, Belopolski, Neupane, Bian, Alidoust, Chang
  et~al.}}]{Xu15sci}
\bibinfo{author}{\bibfnamefont{S.~Y.} \bibnamefont{Xu}},
  \bibinfo{author}{\bibfnamefont{C.}~\bibnamefont{Liu}},
  \bibinfo{author}{\bibfnamefont{S.~K.} \bibnamefont{Kushwaha}},
  \bibinfo{author}{\bibfnamefont{R.}~\bibnamefont{Sankar}},
  \bibinfo{author}{\bibfnamefont{J.~W.} \bibnamefont{Krizan}},
  \bibinfo{author}{\bibfnamefont{I.}~\bibnamefont{Belopolski}},
  \bibinfo{author}{\bibfnamefont{M.}~\bibnamefont{Neupane}},
  \bibinfo{author}{\bibfnamefont{G.}~\bibnamefont{Bian}},
  \bibinfo{author}{\bibfnamefont{N.}~\bibnamefont{Alidoust}},
  \bibinfo{author}{\bibfnamefont{T.~R.} \bibnamefont{Chang}},
  \bibnamefont{et~al.}, \bibinfo{journal}{Science}
  \textbf{\bibinfo{volume}{347}}, \bibinfo{pages}{294}
  (\bibinfo{year}{2015}{\natexlab{b}}),
  \urlprefix\url{http://www.sciencemag.org/content/347/6219/294.abstract}.

\bibitem[{\citenamefont{Burkov et~al.}(2011)\citenamefont{Burkov, Hook, and
  Balents}}]{Burkov11prb}
\bibinfo{author}{\bibfnamefont{A.~A.} \bibnamefont{Burkov}},
  \bibinfo{author}{\bibfnamefont{M.~D.} \bibnamefont{Hook}}, \bibnamefont{and}
  \bibinfo{author}{\bibfnamefont{L.}~\bibnamefont{Balents}},
  \bibinfo{journal}{Phys. Rev. B} \textbf{\bibinfo{volume}{84}},
  \bibinfo{pages}{235126} (\bibinfo{year}{2011}),
  \urlprefix\url{https://link.aps.org/doi/10.1103/PhysRevB.84.235126}.

\bibitem[{\citenamefont{Bian et~al.}(2016)\citenamefont{Bian, Chang, Sankar,
  Xu, Zheng, Neupert, Chiu, Huang, Chang, Belopolski et~al.}}]{Bian16nc}
\bibinfo{author}{\bibfnamefont{G.}~\bibnamefont{Bian}},
  \bibinfo{author}{\bibfnamefont{T.-R.} \bibnamefont{Chang}},
  \bibinfo{author}{\bibfnamefont{R.}~\bibnamefont{Sankar}},
  \bibinfo{author}{\bibfnamefont{S.-Y.} \bibnamefont{Xu}},
  \bibinfo{author}{\bibfnamefont{H.}~\bibnamefont{Zheng}},
  \bibinfo{author}{\bibfnamefont{T.}~\bibnamefont{Neupert}},
  \bibinfo{author}{\bibfnamefont{C.-K.} \bibnamefont{Chiu}},
  \bibinfo{author}{\bibfnamefont{S.-M.} \bibnamefont{Huang}},
  \bibinfo{author}{\bibfnamefont{G.}~\bibnamefont{Chang}},
  \bibinfo{author}{\bibfnamefont{I.}~\bibnamefont{Belopolski}},
  \bibnamefont{et~al.}, \textbf{\bibinfo{volume}{7}}, \bibinfo{pages}{10556}
  (\bibinfo{year}{2016}), \bibinfo{note}{article},
  \urlprefix\url{http://dx.doi.org/10.1038/ncomms10556}.

\bibitem[{\citenamefont{Schoop et~al.}(2016)\citenamefont{Schoop, Ali,
  Stra{\ss}er, Topp, Varykhalov, Marchenko, Duppel, Parkin, Lotsch, and
  Ast}}]{Schoop16nc}
\bibinfo{author}{\bibfnamefont{L.~M.} \bibnamefont{Schoop}},
  \bibinfo{author}{\bibfnamefont{M.~N.} \bibnamefont{Ali}},
  \bibinfo{author}{\bibfnamefont{C.}~\bibnamefont{Stra{\ss}er}},
  \bibinfo{author}{\bibfnamefont{A.}~\bibnamefont{Topp}},
  \bibinfo{author}{\bibfnamefont{A.}~\bibnamefont{Varykhalov}},
  \bibinfo{author}{\bibfnamefont{D.}~\bibnamefont{Marchenko}},
  \bibinfo{author}{\bibfnamefont{V.}~\bibnamefont{Duppel}},
  \bibinfo{author}{\bibfnamefont{S.~S.~P.} \bibnamefont{Parkin}},
  \bibinfo{author}{\bibfnamefont{B.~V.} \bibnamefont{Lotsch}},
  \bibnamefont{and} \bibinfo{author}{\bibfnamefont{C.~R.} \bibnamefont{Ast}},
  \bibinfo{journal}{Nature Commun.} \textbf{\bibinfo{volume}{7}},
  \bibinfo{pages}{11696} (\bibinfo{year}{2016}),
  \urlprefix\url{http://www.nature.com/ncomms/2016/160531/ncomms11696/full/ncomms11696.html}.

\bibitem[{\citenamefont{Neupane et~al.}(2016)\citenamefont{Neupane, Belopolski,
  Hosen, Sanchez, Sankar, Szlawska, Xu, Dimitri, Dhakal, Maldonado
  et~al.}}]{Neupane16prb}
\bibinfo{author}{\bibfnamefont{M.}~\bibnamefont{Neupane}},
  \bibinfo{author}{\bibfnamefont{I.}~\bibnamefont{Belopolski}},
  \bibinfo{author}{\bibfnamefont{M.~M.} \bibnamefont{Hosen}},
  \bibinfo{author}{\bibfnamefont{D.~S.} \bibnamefont{Sanchez}},
  \bibinfo{author}{\bibfnamefont{R.}~\bibnamefont{Sankar}},
  \bibinfo{author}{\bibfnamefont{M.}~\bibnamefont{Szlawska}},
  \bibinfo{author}{\bibfnamefont{S.-Y.} \bibnamefont{Xu}},
  \bibinfo{author}{\bibfnamefont{K.}~\bibnamefont{Dimitri}},
  \bibinfo{author}{\bibfnamefont{N.}~\bibnamefont{Dhakal}},
  \bibinfo{author}{\bibfnamefont{P.}~\bibnamefont{Maldonado}},
  \bibnamefont{et~al.}, \bibinfo{journal}{Phys. Rev. B}
  \textbf{\bibinfo{volume}{93}}, \bibinfo{pages}{201104}
  (\bibinfo{year}{2016}),
  \urlprefix\url{https://journals.aps.org/prb/abstract/10.1103/PhysRevB.93.201104}.

\bibitem[{\citenamefont{Wu et~al.}(2016)\citenamefont{Wu, Wang, Mun, Johnson,
  Mou, Huang, Lee, Bud'ko, Canfield, and Kaminski}}]{Wu16np}
\bibinfo{author}{\bibfnamefont{Y.}~\bibnamefont{Wu}},
  \bibinfo{author}{\bibfnamefont{L.-L.} \bibnamefont{Wang}},
  \bibinfo{author}{\bibfnamefont{E.}~\bibnamefont{Mun}},
  \bibinfo{author}{\bibfnamefont{D.~D.} \bibnamefont{Johnson}},
  \bibinfo{author}{\bibfnamefont{D.}~\bibnamefont{Mou}},
  \bibinfo{author}{\bibfnamefont{L.}~\bibnamefont{Huang}},
  \bibinfo{author}{\bibfnamefont{Y.}~\bibnamefont{Lee}},
  \bibinfo{author}{\bibfnamefont{S.~L.} \bibnamefont{Bud'ko}},
  \bibinfo{author}{\bibfnamefont{P.~C.} \bibnamefont{Canfield}},
  \bibnamefont{and} \bibinfo{author}{\bibfnamefont{A.}~\bibnamefont{Kaminski}},
  \bibinfo{journal}{Nature Phys.} \textbf{\bibinfo{volume}{12}},
  \bibinfo{pages}{667} (\bibinfo{year}{2016}),
  \urlprefix\url{http://www.nature.com/nphys/journal/vaop/ncurrent/full/nphys3712.html?WT.feed_name=subjects_electronic-properties-and-materials}.

\bibitem[{\citenamefont{Wang et~al.}(2016)\citenamefont{Wang, Alexandradinata,
  Cava, and Bernevig}}]{WangZJ16nat}
\bibinfo{author}{\bibfnamefont{Z.}~\bibnamefont{Wang}},
  \bibinfo{author}{\bibfnamefont{A.}~\bibnamefont{Alexandradinata}},
  \bibinfo{author}{\bibfnamefont{R.~J.} \bibnamefont{Cava}}, \bibnamefont{and}
  \bibinfo{author}{\bibfnamefont{B.~A.} \bibnamefont{Bernevig}},
  \bibinfo{journal}{Nature} \textbf{\bibinfo{volume}{532}},
  \bibinfo{pages}{189} (\bibinfo{year}{2016}), ISSN \bibinfo{issn}{0028-0836},
  \bibinfo{note}{article},
  \urlprefix\url{http://dx.doi.org/10.1038/nature17410}.

\bibitem[{\citenamefont{Chen et~al.}(2017)\citenamefont{Chen, Lu, and
  Hou}}]{ChenW17arXiv}
\bibinfo{author}{\bibfnamefont{W.}~\bibnamefont{Chen}},
  \bibinfo{author}{\bibfnamefont{H.-Z.} \bibnamefont{Lu}}, \bibnamefont{and}
  \bibinfo{author}{\bibfnamefont{J.-M.} \bibnamefont{Hou}},
  \bibinfo{journal}{arXiv:1703.10886}  (\bibinfo{year}{2017}),
  \urlprefix\url{http://arxiv.org/abs/1703.10886}.

\bibitem[{\citenamefont{Yan et~al.}(2017)\citenamefont{Yan, Bi, Shen, Lu,
  Zhang, and Wang}}]{YanZ17arXiv}
\bibinfo{author}{\bibfnamefont{Z.}~\bibnamefont{Yan}},
  \bibinfo{author}{\bibfnamefont{R.}~\bibnamefont{Bi}},
  \bibinfo{author}{\bibfnamefont{H.}~\bibnamefont{Shen}},
  \bibinfo{author}{\bibfnamefont{L.}~\bibnamefont{Lu}},
  \bibinfo{author}{\bibfnamefont{S.-C.} \bibnamefont{Zhang}}, \bibnamefont{and}
  \bibinfo{author}{\bibfnamefont{Z.}~\bibnamefont{Wang}},
  \bibinfo{journal}{arXiv:1704.00655}  (\bibinfo{year}{2017}),
  \urlprefix\url{http://arxiv.org/abs/1704.00655}.

\bibitem[{\citenamefont{Zheng and Ando}(2002)}]{ZhengYS02prb}
\bibinfo{author}{\bibfnamefont{Y.}~\bibnamefont{Zheng}} \bibnamefont{and}
  \bibinfo{author}{\bibfnamefont{T.}~\bibnamefont{Ando}},
  \bibinfo{journal}{Phys. Rev. B} \textbf{\bibinfo{volume}{65}},
  \bibinfo{pages}{245420} (\bibinfo{year}{2002}),
  \urlprefix\url{https://link.aps.org/doi/10.1103/PhysRevB.65.245420}.

\bibitem[{\citenamefont{Gusynin and Sharapov}(2005)}]{Gusynin05prl}
\bibinfo{author}{\bibfnamefont{V.~P.} \bibnamefont{Gusynin}} \bibnamefont{and}
  \bibinfo{author}{\bibfnamefont{S.~G.} \bibnamefont{Sharapov}},
  \bibinfo{journal}{Phys. Rev. Lett.} \textbf{\bibinfo{volume}{95}},
  \bibinfo{pages}{146801} (\bibinfo{year}{2005}),
  \urlprefix\url{https://link.aps.org/doi/10.1103/PhysRevLett.95.146801}.

\bibitem[{\citenamefont{Novoselov et~al.}(2005)\citenamefont{Novoselov, Geim,
  Morozov, Jiang, Katsnelson, Grigorieva, Dubonos, and
  Firsov}}]{Novoselov05nat}
\bibinfo{author}{\bibfnamefont{K.~S.} \bibnamefont{Novoselov}},
  \bibinfo{author}{\bibfnamefont{A.~K.} \bibnamefont{Geim}},
  \bibinfo{author}{\bibfnamefont{S.~V.} \bibnamefont{Morozov}},
  \bibinfo{author}{\bibfnamefont{D.}~\bibnamefont{Jiang}},
  \bibinfo{author}{\bibfnamefont{M.~I.} \bibnamefont{Katsnelson}},
  \bibinfo{author}{\bibfnamefont{I.~V.} \bibnamefont{Grigorieva}},
  \bibinfo{author}{\bibfnamefont{S.~V.} \bibnamefont{Dubonos}},
  \bibnamefont{and} \bibinfo{author}{\bibfnamefont{A.~A.}
  \bibnamefont{Firsov}}, \bibinfo{journal}{Nature}
  \textbf{\bibinfo{volume}{438}}, \bibinfo{pages}{197} (\bibinfo{year}{2005}),
  ISSN \bibinfo{issn}{0028-0836},
  \urlprefix\url{http://dx.doi.org/10.1038/nature04233}.

\bibitem[{\citenamefont{Zhang et~al.}(2005)\citenamefont{Zhang, Tan, Stormer,
  and Kim}}]{ZhangYB05nat}
\bibinfo{author}{\bibfnamefont{Y.}~\bibnamefont{Zhang}},
  \bibinfo{author}{\bibfnamefont{Y.-W.} \bibnamefont{Tan}},
  \bibinfo{author}{\bibfnamefont{H.~L.} \bibnamefont{Stormer}},
  \bibnamefont{and} \bibinfo{author}{\bibfnamefont{P.}~\bibnamefont{Kim}},
  \bibinfo{journal}{Nature} \textbf{\bibinfo{volume}{438}},
  \bibinfo{pages}{201} (\bibinfo{year}{2005}),
  \urlprefix\url{http://www.nature.com/nature/journal/v438/n7065/abs/nature04235.html}.

\bibitem[{\citenamefont{Xu et~al.}(2014)\citenamefont{Xu, Miotkowski, Liu,
  Tian, Nam, Alidoust, Hu, Shih, Hasan, and Chen}}]{Xu14np}
\bibinfo{author}{\bibfnamefont{Y.}~\bibnamefont{Xu}},
  \bibinfo{author}{\bibfnamefont{I.}~\bibnamefont{Miotkowski}},
  \bibinfo{author}{\bibfnamefont{C.}~\bibnamefont{Liu}},
  \bibinfo{author}{\bibfnamefont{J.}~\bibnamefont{Tian}},
  \bibinfo{author}{\bibfnamefont{H.}~\bibnamefont{Nam}},
  \bibinfo{author}{\bibfnamefont{N.}~\bibnamefont{Alidoust}},
  \bibinfo{author}{\bibfnamefont{J.}~\bibnamefont{Hu}},
  \bibinfo{author}{\bibfnamefont{C.-K.} \bibnamefont{Shih}},
  \bibinfo{author}{\bibfnamefont{M.~Z.} \bibnamefont{Hasan}}, \bibnamefont{and}
  \bibinfo{author}{\bibfnamefont{Y.~P.} \bibnamefont{Chen}},
  \bibinfo{journal}{Nature Phys.} \textbf{\bibinfo{volume}{10}},
  \bibinfo{pages}{956} (\bibinfo{year}{2014}),
  \urlprefix\url{http://www.nature.com/nphys/journal/v10/n12/full/nphys3140.html}.

\bibitem[{\citenamefont{Yoshimi et~al.}(2015)\citenamefont{Yoshimi, Yasuda,
  Tsukazaki, Takahashi, Nagaosa, Kawasaki, and Tokura}}]{Yoshimi15nc}
\bibinfo{author}{\bibfnamefont{R.}~\bibnamefont{Yoshimi}},
  \bibinfo{author}{\bibfnamefont{K.}~\bibnamefont{Yasuda}},
  \bibinfo{author}{\bibfnamefont{A.}~\bibnamefont{Tsukazaki}},
  \bibinfo{author}{\bibfnamefont{K.~S.} \bibnamefont{Takahashi}},
  \bibinfo{author}{\bibfnamefont{N.}~\bibnamefont{Nagaosa}},
  \bibinfo{author}{\bibfnamefont{M.}~\bibnamefont{Kawasaki}}, \bibnamefont{and}
  \bibinfo{author}{\bibfnamefont{Y.}~\bibnamefont{Tokura}},
  \bibinfo{journal}{Nature Commun.} \textbf{\bibinfo{volume}{6}}
  (\bibinfo{year}{2015}),
  \urlprefix\url{http://www.nature.com/articles/ncomms9530}.

\bibitem[{\citenamefont{Zhang et~al.}(2015{\natexlab{a}})\citenamefont{Zhang,
  Lu, and Shen}}]{ZhangSB15srep}
\bibinfo{author}{\bibfnamefont{S.~B.} \bibnamefont{Zhang}},
  \bibinfo{author}{\bibfnamefont{H.~Z.} \bibnamefont{Lu}}, \bibnamefont{and}
  \bibinfo{author}{\bibfnamefont{S.~Q.} \bibnamefont{Shen}},
  \bibinfo{journal}{Scientific reports} \textbf{\bibinfo{volume}{5}},
  \bibinfo{pages}{13277} (\bibinfo{year}{2015}{\natexlab{a}}),
  \urlprefix\url{http://www.nature.com/articles/srep13277}.

\bibitem[{\citenamefont{Nielsen and Ninomiya}(1983)}]{Nielsen83plb}
\bibinfo{author}{\bibfnamefont{H.~B.} \bibnamefont{Nielsen}} \bibnamefont{and}
  \bibinfo{author}{\bibfnamefont{M.}~\bibnamefont{Ninomiya}},
  \bibinfo{journal}{Physics Letters B} \textbf{\bibinfo{volume}{130}},
  \bibinfo{pages}{389 } (\bibinfo{year}{1983}), ISSN \bibinfo{issn}{0370-2693},
  \urlprefix\url{http://www.sciencedirect.com/science/article/pii/0370269383915290}.

\bibitem[{\citenamefont{Alicea and Balents}(2009)}]{Alicea09prbrc}
\bibinfo{author}{\bibfnamefont{J.}~\bibnamefont{Alicea}} \bibnamefont{and}
  \bibinfo{author}{\bibfnamefont{L.}~\bibnamefont{Balents}},
  \bibinfo{journal}{Phys. Rev. B} \textbf{\bibinfo{volume}{79}},
  \bibinfo{pages}{241101} (\bibinfo{year}{2009}),
  \urlprefix\url{https://link.aps.org/doi/10.1103/PhysRevB.79.241101}.

\bibitem[{\citenamefont{Wang and Zhang}(2013)}]{WangZ13prbrc}
\bibinfo{author}{\bibfnamefont{Z.}~\bibnamefont{Wang}} \bibnamefont{and}
  \bibinfo{author}{\bibfnamefont{S.-C.} \bibnamefont{Zhang}},
  \bibinfo{journal}{Phys. Rev. B} \textbf{\bibinfo{volume}{87}},
  \bibinfo{pages}{161107} (\bibinfo{year}{2013}),
  \urlprefix\url{https://link.aps.org/doi/10.1103/PhysRevB.87.161107}.

\bibitem[{\citenamefont{Zhang and Nagaosa}(2017)}]{ZhangXX17prb}
\bibinfo{author}{\bibfnamefont{X.-X.} \bibnamefont{Zhang}} \bibnamefont{and}
  \bibinfo{author}{\bibfnamefont{N.}~\bibnamefont{Nagaosa}},
  \bibinfo{journal}{Phys. Rev. B} \textbf{\bibinfo{volume}{95}},
  \bibinfo{pages}{205143} (\bibinfo{year}{2017}),
  \urlprefix\url{https://link.aps.org/doi/10.1103/PhysRevB.95.205143}.

\bibitem[{\citenamefont{Zhang et~al.}(2015{\natexlab{b}})\citenamefont{Zhang,
  Lin, Guo, Xu, Lee, Lu, Huang, Chang, Hsu, Lin et~al.}}]{ZhangCL15arXiv-TaP}
\bibinfo{author}{\bibfnamefont{C.}~\bibnamefont{Zhang}},
  \bibinfo{author}{\bibfnamefont{Z.}~\bibnamefont{Lin}},
  \bibinfo{author}{\bibfnamefont{C.}~\bibnamefont{Guo}},
  \bibinfo{author}{\bibfnamefont{S.-Y.} \bibnamefont{Xu}},
  \bibinfo{author}{\bibfnamefont{C.-C.} \bibnamefont{Lee}},
  \bibinfo{author}{\bibfnamefont{H.}~\bibnamefont{Lu}},
  \bibinfo{author}{\bibfnamefont{S.-M.} \bibnamefont{Huang}},
  \bibinfo{author}{\bibfnamefont{G.}~\bibnamefont{Chang}},
  \bibinfo{author}{\bibfnamefont{C.-H.} \bibnamefont{Hsu}},
  \bibinfo{author}{\bibfnamefont{H.}~\bibnamefont{Lin}}, \bibnamefont{et~al.},
  \bibinfo{journal}{arXiv:1507.06301}  (\bibinfo{year}{2015}{\natexlab{b}}),
  \urlprefix\url{http://arxiv.org/abs/1507.06301}.

\bibitem[{\citenamefont{Kitaev}(2006)}]{Kitaev-Honeycoumb}
\bibinfo{author}{\bibfnamefont{A.}~\bibnamefont{Kitaev}},
  \bibinfo{journal}{Annals of Physics} \textbf{\bibinfo{volume}{321}},
  \bibinfo{pages}{2 } (\bibinfo{year}{2006}), ISSN \bibinfo{issn}{0003-4916},
  \bibinfo{note}{january Special Issue},
  \urlprefix\url{http://www.sciencedirect.com/science/article/pii/S0003491605002381}.

\bibitem[{\citenamefont{Hastings and Loring}(2010)}]{Hastings1}
\bibinfo{author}{\bibfnamefont{M.~B.} \bibnamefont{Hastings}} \bibnamefont{and}
  \bibinfo{author}{\bibfnamefont{T.~A.} \bibnamefont{Loring}},
  \bibinfo{journal}{Journal of Mathematical Physics}
  \textbf{\bibinfo{volume}{51}}, \bibinfo{pages}{015214}
  (\bibinfo{year}{2010}), \urlprefix\url{http://dx.doi.org/10.1063/1.3274817}.

\bibitem[{\citenamefont{Hastings and Loring}(2011)}]{Hastings2}
\bibinfo{author}{\bibfnamefont{M.~B.} \bibnamefont{Hastings}} \bibnamefont{and}
  \bibinfo{author}{\bibfnamefont{T.~A.} \bibnamefont{Loring}},
  \bibinfo{journal}{Annals of Physics} \textbf{\bibinfo{volume}{326}},
  \bibinfo{pages}{1699 } (\bibinfo{year}{2011}), ISSN
  \bibinfo{issn}{0003-4916}, \bibinfo{note}{july 2011 Special Issue},
  \urlprefix\url{http://www.sciencedirect.com/science/article/pii/S0003491610002277}.

\bibitem[{\citenamefont{Prodan and Schulz-Baldes}(2016)}]{Prodan-book}
\bibinfo{author}{\bibfnamefont{E.}~\bibnamefont{Prodan}} \bibnamefont{and}
  \bibinfo{author}{\bibfnamefont{H.}~\bibnamefont{Schulz-Baldes}},
  \emph{\bibinfo{title}{Bulk and Boundary Invariants for Complex Topological
  Insulators: From K-Theory to Physics}} (\bibinfo{publisher}{Springer
  International Publishing, Switzerland}, \bibinfo{year}{2016}).

\bibitem[{\citenamefont{Hasan and Kane}(2010{\natexlab{b}})}]{Kane-RMP}
\bibinfo{author}{\bibfnamefont{M.~Z.} \bibnamefont{Hasan}} \bibnamefont{and}
  \bibinfo{author}{\bibfnamefont{C.~L.} \bibnamefont{Kane}},
  \bibinfo{journal}{Rev. Mod. Phys.} \textbf{\bibinfo{volume}{82}},
  \bibinfo{pages}{3045} (\bibinfo{year}{2010}{\natexlab{b}}),
  \urlprefix\url{http://link.aps.org/doi/10.1103/RevModPhys.82.3045}.

\bibitem[{\citenamefont{Qi and Zhang}(2011{\natexlab{b}})}]{XLQi-RMP}
\bibinfo{author}{\bibfnamefont{X.-L.} \bibnamefont{Qi}} \bibnamefont{and}
  \bibinfo{author}{\bibfnamefont{S.-C.} \bibnamefont{Zhang}},
  \bibinfo{journal}{Rev. Mod. Phys.} \textbf{\bibinfo{volume}{83}},
  \bibinfo{pages}{1057} (\bibinfo{year}{2011}{\natexlab{b}}),
  \urlprefix\url{http://link.aps.org/doi/10.1103/RevModPhys.83.1057}.

\bibitem[{\citenamefont{Schnyder et~al.}(2008)\citenamefont{Schnyder, Ryu,
  Furusaki, and Ludwig}}]{Schnyder-classification}
\bibinfo{author}{\bibfnamefont{A.~P.} \bibnamefont{Schnyder}},
  \bibinfo{author}{\bibfnamefont{S.}~\bibnamefont{Ryu}},
  \bibinfo{author}{\bibfnamefont{A.}~\bibnamefont{Furusaki}}, \bibnamefont{and}
  \bibinfo{author}{\bibfnamefont{A.~W.~W.} \bibnamefont{Ludwig}},
  \bibinfo{journal}{Phys. Rev. B} \textbf{\bibinfo{volume}{78}},
  \bibinfo{pages}{195125} (\bibinfo{year}{2008}),
  \urlprefix\url{http://link.aps.org/doi/10.1103/PhysRevB.78.195125}.

\bibitem[{\citenamefont{Kitaev}(2009)}]{Kitaev-classification}
\bibinfo{author}{\bibfnamefont{A.}~\bibnamefont{Kitaev}}, \bibinfo{journal}{AIP
  Conference Proceedings} \textbf{\bibinfo{volume}{1134}}, \bibinfo{pages}{22}
  (\bibinfo{year}{2009}),
  \urlprefix\url{http://aip.scitation.org/doi/abs/10.1063/1.3149495}.

\bibitem[{\citenamefont{Ryu et~al.}(2010)\citenamefont{Ryu, Schnyder, Furusaki,
  and Ludwig}}]{Ryu-classification}
\bibinfo{author}{\bibfnamefont{S.}~\bibnamefont{Ryu}},
  \bibinfo{author}{\bibfnamefont{A.~P.} \bibnamefont{Schnyder}},
  \bibinfo{author}{\bibfnamefont{A.}~\bibnamefont{Furusaki}}, \bibnamefont{and}
  \bibinfo{author}{\bibfnamefont{A.~W.~W.} \bibnamefont{Ludwig}},
  \bibinfo{journal}{New Journal of Physics} \textbf{\bibinfo{volume}{12}},
  \bibinfo{pages}{065010} (\bibinfo{year}{2010}),
  \urlprefix\url{http://stacks.iop.org/1367-2630/12/i=6/a=065010}.

\bibitem[{\citenamefont{Hatsugai}(1993)}]{Hasugai-Chern}
\bibinfo{author}{\bibfnamefont{Y.}~\bibnamefont{Hatsugai}},
  \bibinfo{journal}{Phys. Rev. Lett.} \textbf{\bibinfo{volume}{71}},
  \bibinfo{pages}{3697} (\bibinfo{year}{1993}),
  \urlprefix\url{https://link.aps.org/doi/10.1103/PhysRevLett.71.3697}.

\bibitem[{\citenamefont{Zhao and Wang}(2014)}]{ZhaoWang-BBC}
\bibinfo{author}{\bibfnamefont{Y.~X.} \bibnamefont{Zhao}} \bibnamefont{and}
  \bibinfo{author}{\bibfnamefont{Z.~D.} \bibnamefont{Wang}},
  \bibinfo{journal}{Phys. Rev. B} \textbf{\bibinfo{volume}{89}},
  \bibinfo{pages}{075111} (\bibinfo{year}{2014}),
  \urlprefix\url{http://link.aps.org/doi/10.1103/PhysRevB.89.075111}.

\bibitem[{Not()}]{Notes-topological-number}
\bibinfo{note}{In numerical computing, edges with width of the order of
  tunnelling distance are cut off from the matrices in the trace to avoid
  boundary effects.}

\bibitem[{\citenamefont{Su et~al.}(1979)\citenamefont{Su, Schrieffer, and
  Heeger}}]{SSH-Model-1}
\bibinfo{author}{\bibfnamefont{W.~P.} \bibnamefont{Su}},
  \bibinfo{author}{\bibfnamefont{J.~R.} \bibnamefont{Schrieffer}},
  \bibnamefont{and} \bibinfo{author}{\bibfnamefont{A.~J.}
  \bibnamefont{Heeger}}, \bibinfo{journal}{Phys. Rev. Lett.}
  \textbf{\bibinfo{volume}{42}}, \bibinfo{pages}{1698} (\bibinfo{year}{1979}),
  \urlprefix\url{https://link.aps.org/doi/10.1103/PhysRevLett.42.1698}.

\bibitem[{\citenamefont{Su et~al.}(1980)\citenamefont{Su, Schrieffer, and
  Heeger}}]{SSH-Model-2}
\bibinfo{author}{\bibfnamefont{W.~P.} \bibnamefont{Su}},
  \bibinfo{author}{\bibfnamefont{J.~R.} \bibnamefont{Schrieffer}},
  \bibnamefont{and} \bibinfo{author}{\bibfnamefont{A.~J.}
  \bibnamefont{Heeger}}, \bibinfo{journal}{Phys. Rev. B}
  \textbf{\bibinfo{volume}{22}}, \bibinfo{pages}{2099} (\bibinfo{year}{1980}),
  \urlprefix\url{https://link.aps.org/doi/10.1103/PhysRevB.22.2099}.

\bibitem[{Sup()}]{Supp}
\bibinfo{note}{See Supplemental Materials for....}

\bibitem[{\citenamefont{Haldane}(1988)}]{Haldane-Model}
\bibinfo{author}{\bibfnamefont{F.~D.~M.} \bibnamefont{Haldane}},
  \bibinfo{journal}{Phys. Rev. Lett.} \textbf{\bibinfo{volume}{61}},
  \bibinfo{pages}{2015} (\bibinfo{year}{1988}),
  \urlprefix\url{http://link.aps.org/doi/10.1103/PhysRevLett.61.2015}.

\bibitem[{\citenamefont{Thouless
  et~al.}(1982{\natexlab{b}})\citenamefont{Thouless, Kohmoto, Nightingale, and
  den Nijs}}]{TKNN}
\bibinfo{author}{\bibfnamefont{D.~J.} \bibnamefont{Thouless}},
  \bibinfo{author}{\bibfnamefont{M.}~\bibnamefont{Kohmoto}},
  \bibinfo{author}{\bibfnamefont{M.~P.} \bibnamefont{Nightingale}},
  \bibnamefont{and} \bibinfo{author}{\bibfnamefont{M.}~\bibnamefont{den Nijs}},
  \bibinfo{journal}{Phys. Rev. Lett.} \textbf{\bibinfo{volume}{49}},
  \bibinfo{pages}{405} (\bibinfo{year}{1982}{\natexlab{b}}),
  \urlprefix\url{http://link.aps.org/doi/10.1103/PhysRevLett.49.405}.

\bibitem[{\citenamefont{Nielsen and Ninomiya}(1981)}]{NN-NoGo}
\bibinfo{author}{\bibfnamefont{H.}~\bibnamefont{Nielsen}} \bibnamefont{and}
  \bibinfo{author}{\bibfnamefont{M.}~\bibnamefont{Ninomiya}},
  \bibinfo{journal}{Nucl. Phys. B} \textbf{\bibinfo{volume}{185}},
  \bibinfo{pages}{20 } (\bibinfo{year}{1981}),
  \urlprefix\url{http://www.sciencedirect.com/science/article/pii/0550321381903618}.

\bibitem[{\citenamefont{Zhao and Wang}(2016)}]{ZhaoWang-NoGo}
\bibinfo{author}{\bibfnamefont{Y.~X.} \bibnamefont{Zhao}} \bibnamefont{and}
  \bibinfo{author}{\bibfnamefont{Z.~D.} \bibnamefont{Wang}},
  \bibinfo{journal}{Phys. Rev. Lett.} \textbf{\bibinfo{volume}{116}},
  \bibinfo{pages}{016401} (\bibinfo{year}{2016}),
  \urlprefix\url{http://link.aps.org/doi/10.1103/PhysRevLett.116.016401}.

\bibitem[{\citenamefont{Volovik}(2011)}]{Volovik11arXiv}
\bibinfo{author}{\bibfnamefont{G.~E.} \bibnamefont{Volovik}},
  \bibinfo{journal}{arXiv:1111.4627}  (\bibinfo{year}{2011}),
  \urlprefix\url{https://arxiv.org/abs/1111.4627}.

\bibitem[{\citenamefont{Lu and Shen}(2017)}]{Lu17fop}
\bibinfo{author}{\bibfnamefont{H.-Z.} \bibnamefont{Lu}} \bibnamefont{and}
  \bibinfo{author}{\bibfnamefont{S.-Q.} \bibnamefont{Shen}},
  \bibinfo{journal}{Front. Phys.} \textbf{\bibinfo{volume}{12}},
  \bibinfo{pages}{127201} (\bibinfo{year}{2017}),
  \urlprefix\url{http://link.springer.com/article/10.1007%2Fs11467-016-0609-y}.

\end{thebibliography}

\clearpage
\newpage

\appendix

\renewcommand{\thefigure}{S\arabic{figure}}
\renewcommand{\theequation}{S\arabic{equation}}
\renewcommand{\thesection}{S\arabic{section}}

\begin{center}
	\textbf{
		\large{Supplemental Material}
	}
\end{center}

\vspace{-0.2cm}

\section{2D Dirac fermions with muptile topological charges}
Let us consider 2D massless Dirac fermions with quadratic band touching \cite{Volovik11arXiv},
\begin{equation}
\begin{split}
\mathcal{H}_{D}^{(2)}(\mathbf{p}) &=w~(p_x^2-p_y^2)\sigma_1+2w~p_xp_y\sigma_2\\
&=\frac{w}{2}(p_{-}^2\sigma^++p_+^2\sigma^-).
\end{split}
\end{equation}
The model has mirror reflection symmetries through the $x$ and $y$-axis, corresponding to the operators $\hat{M}_x=i\sigma_1\hat{R}_y$ and $\hat{M}_y=i\sigma_1\hat{R}_x$, respectively, with $\hat{R}_y$ ($\hat{R}_x$) being the inversion of the $y$-coordinate ($x$-coordinate), and therefore has inversion symmetry as the combination of $\hat{M}_x$ and $\hat{M}_y$, $\hat{I}=\hat{M}_x\hat{M}_y$. The model may appear in solid-state systems that satisfy these crystal symmetries.
In a $z$-direction perpendicular magnetic field, the Hamiltonian becomes
\begin{equation}
\mathcal{H}^{(2)}_{D,B}=\frac{\lambda}{2}[(a^\dagger)^2\sigma^++a^2\sigma^-],
\end{equation}
with $\lambda=w(\sqrt{2}\hbar/\ell_B)^2=2w\hbar qB$. The corresponding infinite lattice model is
\begin{equation}
\begin{split}
\mathcal{H}^{(2)}_{D,L}=&\frac{\lambda}{2}(S^\dagger\sqrt{|X|+1}S^\dagger\sqrt{|X|+1}\sigma^+\\
&~~~+\sqrt{|X|+1}S\sqrt{|X|+1}S\sigma^-),
\end{split}
\end{equation}
which gives $Q=\lambda \sqrt{|X|+1}S\sqrt{|X|+1}S$.
The model has an energy gap since $Q$ is invertible with the inverse,
\begin{eqnarray}
Q^{-1}=\frac{1}{\lambda}S^\dagger\frac{1}{\sqrt{|X|+1}}S^\dagger\frac{1}{\sqrt{|X|+1}},
\end{eqnarray}
and can be continuously deformed to be $\tilde{Q}=S^2$ via $Q(\tau)=\lambda \sqrt{|X|e^{-\tau}+1}S\sqrt{|X|e^{-\tau}+1}S$ with $\tau\in[0,\infty)$, during which the gap is opened and the invertibility of $Q$ is preserved. Now we calculate the topological invariant of $\tilde{Q}=S^2$ in momentum space by using the acquired translation invariance, which turns out to be $2$. Thus there exist two zero-energy end-states with positive chirality for the semi-infinite lattice with sites on the left of the 0th site being cut off. Accordingly, there are two-fold degenerate zero-energy Landau levels with positive chirality and small Landau indices.

This model can be generalized to be $\mathcal{H}^{(k)}_D=(w/2)(p_{-}^k\sigma^++p^k_+\sigma^-)$ for any positive integer $k$, which corresponds to a topological invariant $\nu=k$, implying that the Landau levels with positive chirality at zero energy are $k$-fold degenerate besides the Landau degeneracy. To construct models for $\nu=-k$, we just need to exchange $p_-$ and $p_+$, which leads to $\mathcal{H}^{(-k)}_D=(w/2)(p_{+}^k\sigma^++p^k_-\sigma^-)$. Correspondingly for this model there exist zero-energy end-states or equivalently zero-energy Landau levels with small indices, which have negative chirality.
\begin{figure}
	\includegraphics[scale=0.7]{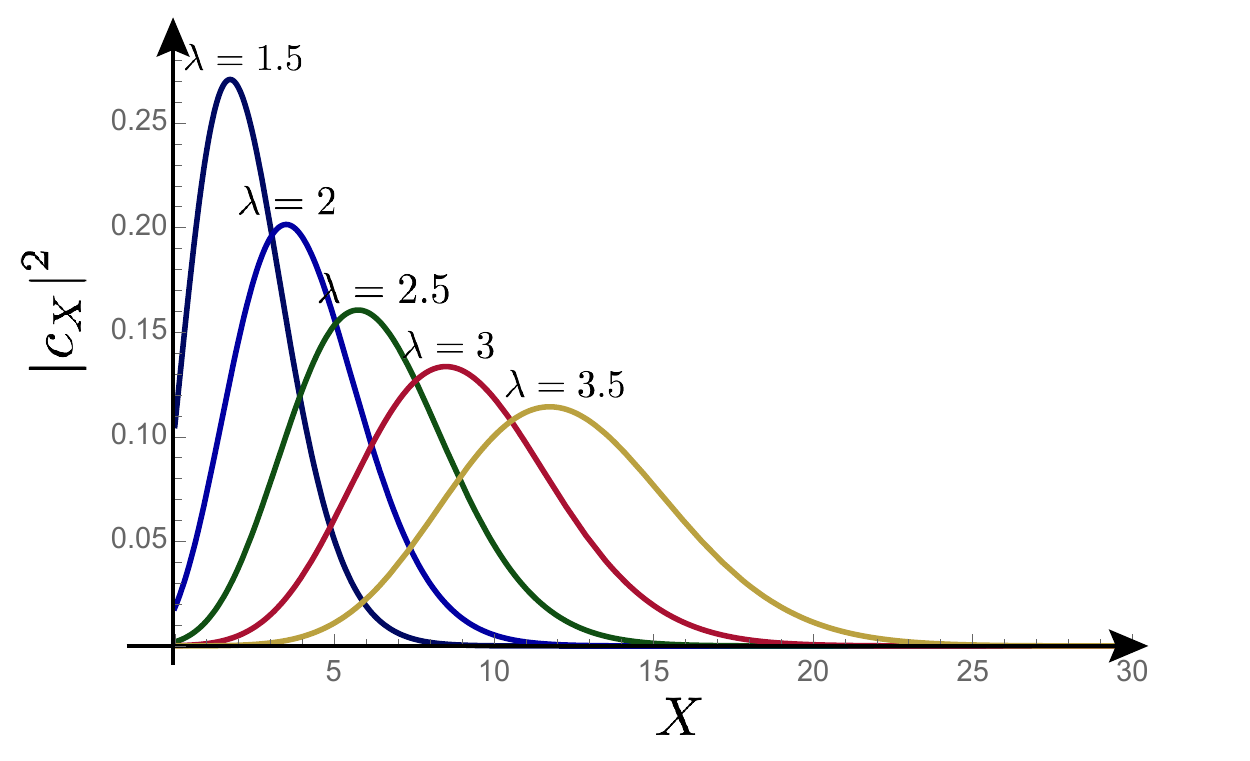}
	\caption{Spatial distributions of zero-energy end-modes for the SSH states with $\lambda=1.5$, $2$, $2.5$, $3$ and $3.5$, respectively. The peak becomes lower and wider with increasing $\lambda$. \label{Edge-modes}}
\end{figure}
\section{Spatial Distributions of Zero-Energy Landau Levels corresponding to generic SSH models}
The zero-energy wave function of $\mathcal{H}_D=(1/2)(a^\dagger\sigma^++a\sigma^-)+\lambda\sigma_2$, or the zero-energy end-state of the equivalent semi-infinite Landau lattice, $\mathcal{H}_D^L=(1/2)(\widehat{S}^\dagger\sqrt{X+1}+\sqrt{X+1}\widehat{S})+\lambda\sigma_2$, can be exactly solved as
\begin{equation}
|\psi\rangle=e^{-\frac{\lambda^2}{2}}\sum_{n=0}^{\infty}\frac{\lambda^n}{\sqrt{n!}}\,|n,\uparrow\rangle.
\end{equation}
The zero-energy state has been normalized for any $\lambda$, and has positive chirality for the chiral symmetry $\sigma_3$, which are consistent with the fact that the topological invariant is equal to $1$ for any $\lambda$. The probability at position $n$ is given by $|c_X|^2=|\langle n|\psi\rangle|^2=e^{-\lambda^2}\lambda^{2n}/n!$, which is plotted with different $\lambda$'s in Fig.~\ref{Edge-modes}. It is observed that the concentration of the wave pocket is moved towards the bulk and its shape is smoothened, when $\lambda$ is increased, although it is always bounded to the end for any $\lambda$.

\begin{figure}
	\includegraphics[scale=0.45]{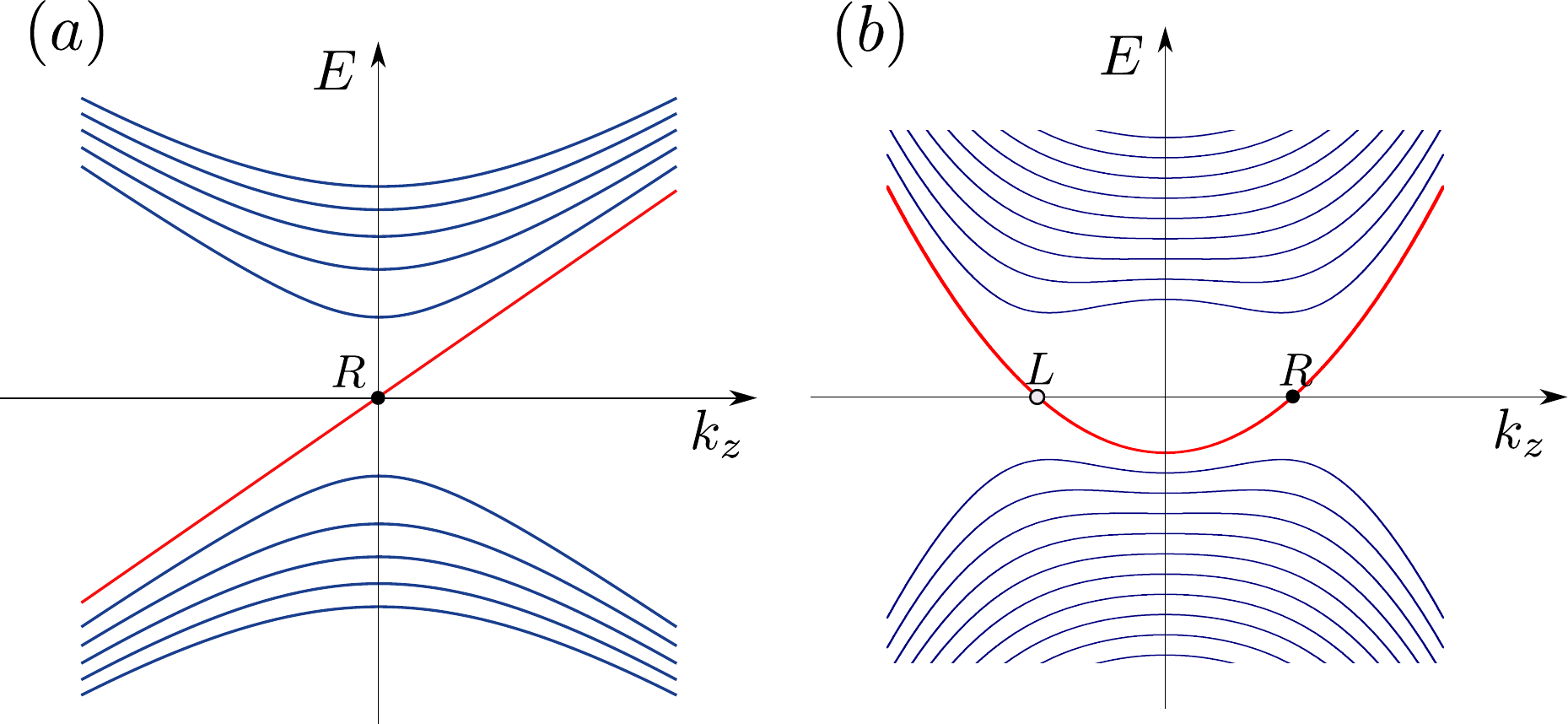}
	\caption{Landau bands for (a) a right-handed Weyl point and (b) a minimal model of Weyl semimetals.  \label{Landau-Weyl}}
\end{figure}
\section{Topological Landau bands of Weyl fermions and Semimetals}
As shown in Fig.~\ref{Landau-Weyl}(a), $\mathcal{H}^R_W=p_x\sigma_1+p_y\sigma_2-p_z\sigma_3$ in a magnetic field has a right-handed chiral band which crosses zero energy and has the smallest Landau index, or equivalently the corresponding Landau lattice is a semi-infinite Chern insulator that has a right-handed chiral edge band. The Landau-band structure for $\mathcal{H}^L_W$ is just the mirror image of Fig.~\ref{Landau-Weyl}(a) through the $E$-axis.

For the minimal model of Weyl semimetals \cite{Lu17fop}, $\mathcal{H}_{WSM}=p_x\sigma_1+p_y\sigma_2-(p_z^2-\mu^2)\sigma_3$, the Landau-band structure is illustrated in Fig.~\ref{Landau-Weyl}(b). The low-energy expansion of $\mathcal{H}_{WSM}$ consists of $\mathcal{H}^R_W$ and $\mathcal{H}^L_W$ at $k_z=\mu$ and $k_z=-\mu$, respectively, and thus the lowest Landau band equivalent to the edge states of the corresponding semi-infinite Landau lattice is obtained by connecting a left-handed chiral band crossing $k_z=-\mu$ and a righ-handed one crossing $k_z=\mu$. The other bands in Fig.~\ref{Landau-Weyl}(b) correspond to the gapped bulk bands of the semi-infinite Landau lattice.

The construction of a Weyl point with a Chern number of $k$ can be achieved by simply adding $-k_z\sigma_3$ to $\mathcal{H}_D^{(k)}$, i.e., $\mathcal{H}_W^{(k)}=\mathcal{H}_D^{(k)}-k_z\sigma_3$. The corresponding semi-infinite Landau lattice is a Chern insulator with a Chern number of $k$, and has  $k$ ($-k$) flavors of right (left)-handed chiral bands on the edge when $k>0$ ($k<0$), or equivelently $k$ ($-k$) flavors of right (left)-handed chiral Landau bands with small indices crossing zero energy.


\end{document}